\documentclass[review]{revtex4}
\usepackage{lineno,hyperref}
\usepackage{amsmath}
\usepackage{amssymb}
\usepackage{graphicx}% Include  files
\usepackage{float}
\modulolinenumbers[10]
\usepackage{mathtools}
\usepackage{xcolor}
\usepackage [autostyle, english = american]{csquotes}
\MakeOuterQuote{"}

\usepackage{xpatch}
\makeatletter
\xpatchcmd{\maketitle}{\if@twoside\next@tpage}{\iffalse}{}{}
\makeatother

\begin{document}

\preprint{APS/123-QED}

\title{Dynamical behavior of the Universe: an entropic force scenario}% Force line breaks with \\
%\thanks{A footnote to the article title}%

\author{Maryam Aghaei Abchouyeh$^{1}$}
   \email{m.aghaei@ph.iut.ac.ir}
   \author{Behrouz Mirza$^{1}$   }
  \email{b.mirza@iut.ac.ir}
   \author{Fatemeh Sadeghi$^{1}$}
   %\email{}

   \affiliation{$^{1}$ Department of Physics, Isfahan University of Technology, Isfahan 84156-83111, Iran}

%\collaboration{CLEO Collaboration}%\noaffiliation

%\date{\today}% It is always \today, today,
             %  but any date may be explicitly specified

%\title{Anyon Black Holes}

%% Group authors per affiliation:
%\author[mymainaddress,mysecondaryaddress]{Maryam Aghaei Abchouyeh\corref{mycorrespondingauthor}}
%\address{Department of Physics, Isfahan University of Technology, Isfahan 84156-83111, Iran}
%\cortext[mycorrespondingauthor]{Corresponding author}
%\fntext[myfootnote]{m.aghaei@ph.iut.ac.ir}
%\ead{m.aghaei@ph.iut.ac.ir}

%% or include affiliations in footnotes:
%\author[mymainaddress]{Behrouz Mirza}
%\ead{b.mirza@cc.iut.ac.ir}

%\author[mymainaddress]{Moein Karimi Takrami}
%\author[mymainaddress]{Younes Younesizadeh}

%\address[mymainaddress]{Department of Physics, Isfahan University of Technology, Isfahan 84156-83111, Iran}
%\address[mysecondaryaddress]{Research Institute for Astronomy and Astrophysics of Maragha (RIAAM) – Maragha, IRAN, P. O. Box: 55134 - 441}

\begin{abstract}
Entropic force originates in the assumption that there is a horizon for the universe. This horizon gives rise to additional terms in the equations of motion. Using dynamical system calculations, our results show that in the presence of dark energy for certain conditions, the last attractor of this theory will be dark energy epoch, but in the absence of dark energy, entropic force energy portion will have the lead role in the late time universe and is responsible for accelerated expansion of that. Interestingly, assuming both entropic force terms and dark energy to have their share of energy density of the universe, we have found that, in certain conditions entropic force dominated epoch is a stable fixed point while the dark energy epoch is a saddle point.

\end{abstract}

%\pacs{Valid PACS appear here}% PACS, the Physics and Astronomy
                             % Classification Scheme.
 %\keywords{Dynamical system, Entropic force, Stability}
%\keywords{Anyon, Van der Waals Black holes, Intermediate statistics}%Use showkeys class option if keyword
                              %display desired
\maketitle

%%%%%%%%%%%%%%%%%%%%%%%%%%%%%%%%%%%%%%%%%%%%%%
\section{Introduction}
%%%%%%%%%%%%%%%%%%%%%%%%%%%%%%%%%%%%%%%%%%%%%%%%	
During the last years of 20th century, the observations proved that the universe is experiencing an accelerated expansion phase \cite{9812133,9805201}.\textcolor{black}{ This expansion can be explained by assuming the cosmological constant which leads to the $\Lambda$CDM model. However, observational data shows that the cosmological constant is around 120 orders of magnitude smaller than its value if we assume the quantum gravity effects to appear at Planck scale, and this would lead to cosmological constant problem. \cite{weinberg,Husain}. There are also other suggestions such as dark energy to explain the accelerated expansion of the universe. Thus the concept of dark energy was introduced as a dynamical field to explain the late time accelerated expansion of the universe \cite{0803.0982}.} Having negative pressure is the main feature of dark energy (as well as cosmological constant), concluding that it has negative equation of state ($\frac{p}{\rho}<0$). But due to ambiguities about dark energy, there are many efforts to avoid that. One of the most popular alternatives to dark energy are modified gravity theories. There are many modified gravitational theories that each of them introduced for a different reason \cite{0705.1032,0805.1726,1104.2669}, although some of them are considered in the presence of dark energy. \\

\indent \textcolor{black}{It is also possible to view the universe using another approach in which the entropic force has the main role. Entropy in macroscopic systems has a tendency to increase which leads to entropic force in many body systems.} \textcolor{black}{If we consider the universe to have a horizon (boundary screen), according to holographic principal the information content of the universe will be encoded on the horizon. Thus the horizon will have the corresponding temperature and entropy. Verlinde used this holographic pirincipal and the Unrah temperature to explain the gravitational interactions as a result of entropic force \cite{verlinde,unrah}.} \textcolor{black}{ Therefore For a homogeneous and isotropic universe with Friedmann-Robertson-Walker ($FRW$) metric, and in natural units ($c=1$),} entropic force interpretation of gravitational interactions, leads to additional terms in the Friedmann equations as bellow  \cite{1002.4278,1003.1528}:
\begin{eqnarray}
\label{eq01}
&& H^2=\dfrac{k}{3}\rho+c_1H^2+c_2\dot{H},\\
\label{eq02}
&& \dot{H}+H^2=\dfrac{-k}{6}(\rho+3p)+c_1H^2+c_2\dot{H},
\end{eqnarray} 
\noindent where, $\rho$ and $p$ are the total energy density and pressure of the universe, $k=8\pi G$ and $H=\frac{\dot{a}}{a}$ is the Hubble parameter. The coefficients $c_1$ and $c_2$ are positive and less than one. Different features of entropic force universe has been studied in literature \cite{1003.4526,1005.0790,1005.1445,1005.2240,1309.7827} and here we are going to study the dynamical behavior of the entropic accelerated universe. In standard cosmology and according to observational information, there is a determined time ordering for the cosmological epochs and their stabilities as bellow:
\begin{equation*}
inflation \rightarrow radiation \rightarrow matter \rightarrow dark\ energy.
\end{equation*} 
\indent A successful gravitational theory must keep these stabilities and ordering. Radiation, matter, curvature and dark energy are the energy \textcolor{black}{components} of the universe which are changing with time, thus a strong tool to investigate if a theory can satisfy this expectation is to use the dynamical system calculations \cite{1409.5585,1512.09281}. Therefore, in the next sections we will study the dynamical behavior of the entropic force universe.\\
\indent This paper is organized as follows: In Sec. (\ref{Sec:2}) the dynamical properties of the entropic force universe in the presence of dark energy is studied and the fixed points of the model and their stabilities are calculated. Sec. \ref{Sec:3} will present a universe in which there is no dark energy and the entropic force will be responsible for late time accelerated expansion of the universe. Finally in Sec. (\ref{Sec:4}) we will consider the universe to include both entropic force and dark energy as the energy \textcolor{black}{components} of the universe and will investigate the stability of dark energy and entropic force energy portions.

%%%%%%%%%%%%%%%%%%%%%%%%%%%%%%%%%
\section{Dark energy existence}
\label{Sec:2}
%%%%%%%%%%%%%%%%%%%%%%%%%%%%%%%%%%%
In this section, we assume \textcolor{black}{dark energy, matter and radiation to be the energy components of the universe,} and the entropic force terms will be considered as usual geometrical terms. Thus the equations of motion will be as follows \cite{1002.4278,1003.1528}:

\begin{eqnarray}
\label{eq1}
&&H^2=\dfrac{k}{3}(\rho_r+\rho_m+\rho_d)+c_1H^2+c_2\dot{H},\\
\label{eq2}
&&\dot{H}+H^2=\dfrac{-k}{6}(\rho_r+\rho_m+\rho_d+3(p_r+p_m+p_d))+c_1H^2+c_2\dot{H},
\end{eqnarray}

\noindent where $\rho_r$, $\rho_m$ and $\rho_d$ are representing the energy density of radiation, matter and dark energy, respectively ($\rho=\rho_r+\rho_m+\rho_d$) and the continuity equations will be written as follows:
\begin{eqnarray}
\label{eq4}
&&\dot{\rho}_m+3H(\rho_m+p_m)=0,\\
&&\dot{\rho}_r+3H(\rho_r+p_r)=0,\\
&&\dot{\rho}_d+3H(\rho_d+p_d)=0,
\end{eqnarray}

\noindent where $p_r=\frac{1}{3}\rho_r$, $p_m=0$ and $p_d=-\rho_d$ are the corresponding pressures for radiation, matter and dark energy. Here we study the case with the equation of state of dark energy to be similar to that of cosmological constant but more complicated cases can be studied using the same method.\\
\indent The above equations implies that our universe can be considered as a dynamical system because its energy \textcolor{black}{components} are changing with respect to time. To study the dynamics of the universe which is described by Eqs.(\ref{eq1}) and (\ref{eq2}), first it is needed to define dimensionless parameters out of the energy densities. Thus we write the energy portions as bellow:
\begin{eqnarray}
\label{eq3}
\Omega_m=\dfrac{k\rho_m}{3H^2},\ \Omega_r=\dfrac{k\rho_r}{3H^2},\ \Omega_d=\dfrac{k\rho_d}{3H^2}.
\end{eqnarray}

Using these definitions, Eqs.\eqref{eq1} and \eqref{eq2} are re-written as a function of energy portions:
\begin{eqnarray}
\label{eq5}
&&1=\Omega_r+\Omega_m+\Omega_d+c_1+c_2\dfrac{\dot{H}}{H^2},\\
\label{eq6}
&&\dfrac{\dot{H}}{H^2}+1=\dfrac{-1}{2}(\Omega_m+2\Omega_r-2\Omega_d)+c_1+c_2\dfrac{\dot{H}}{H^2}.
\end{eqnarray}

To have an absolute expression of $\frac{\dot{H}}{H^2}$ in terms of $\Omega_r$, $\Omega_m$ and $\Omega_d$, the above equations are solved together and the results are as bellow:
\begin{eqnarray}
\label{eq7}
&&1=-\frac{-3 c_2\Omega_m-4 c_2 \Omega_r+2 \Omega_d+2 \Omega_m+2 \Omega_r}{2 (c_1-1)},\\
\label{eq8}
&&\dfrac{\dot{H}}{H^2}=\frac{1}{2} (-3 \Omega_m-4 \Omega_r).
\end{eqnarray}

We are going to use Eqs. \eqref{eq3}, \eqref{eq7} and \eqref{eq8} to calculate the dynamical behavior of the universe. At first the time derivatives of $\Omega_r$, $\Omega_m$ and $\Omega_d$ are calculated, but at the end we made a change of variable from time to number of e-folding "N". $N=Log(a)$ is a measure of universe scale and is more useful than cosmological time. Thus the equations that describe the evolution of the energy portions are as follows:
\begin{eqnarray}
\label{eq9}
\Omega'_r&=&-4\Omega_r-2\Omega_r(\dfrac{\dot{H}}{H^2}),\\
\label{eq10}
\Omega'_m&=&-3\Omega_m-2\Omega_m(\dfrac{\dot{H}}{H^2}),\\
\label{eq11}
\Omega'_d&=&-2\Omega_d(\dfrac{\dot{H}}{H^2}),
\end{eqnarray}

\noindent with prime denoting the derivative with respect to $N$. As Eq.\eqref{eq7} is a conditional term between the energy portions, one of the above equations will be omitted. Here we will omit $\Omega_m$ using Eq.\eqref{eq7}, in order to have two independent equations. The remaining equations should be solved together ($\Omega'_r=0$ and $ \Omega'_d=0$) so that we can find the fixed points of this case. The fixed points of a special case shows the values of each energy portions in which the whole system can be described with a unique behavior. Thus Eqs. \eqref{eq9} and \eqref{eq11} are solved together and the following fixed points are obtained:
\begin{eqnarray}
\label{eq121}
&&1.\ (\Omega_r,\Omega_d)=( 3 c_1-6 c_2+1,\ 0),\nonumber\\
&&2.\ (\Omega_r,\Omega_d)=(0,\  0),\nonumber\\
&&3.\ (\Omega_r,\Omega_d)=(0,\ 1-c_1).
\end{eqnarray}

Now we are at the point to check the stability of these fixed points. The stability of a fixed point can have three different situations. If the flux lines are attracted toward the point, it is called a stable fixed point, if the flux lines are attracted to the fixed point from one direction and leaves the point from another one, it is called a saddle point and if the point acts as a repulsive one, it is unstable. These properties are investigated using the eigenvalues of Jacobian matrix. If we write Eqs.\eqref{eq9} and \eqref{eq11} as:
\begin{eqnarray}
\label{eq13}
\Omega'_r=f(\Omega_r,\Omega_d),\\
\Omega'_d=g(\Omega_r,\Omega_d),
\end{eqnarray}

\noindent the Jacobian matrix is defined as bellow:
\[ J= \left[
\begin{matrix}
\dfrac{\partial f}{\partial \Omega_r} & \dfrac{\partial f}{\partial \Omega_d}\\
\dfrac{\partial g}{\partial \Omega_r}&\dfrac{\partial g}{\partial \Omega_d}\\
\end{matrix}
\right].\]

The eigenvalues of this matrix determines the stability of a fixed point. As our system is a two dimensional dynamical system, Jacobian matrix will have two eigenvalues for each of the fixed points. Therefore:

\begin{figure} % figure 7
	\center
	\includegraphics[scale=0.5]{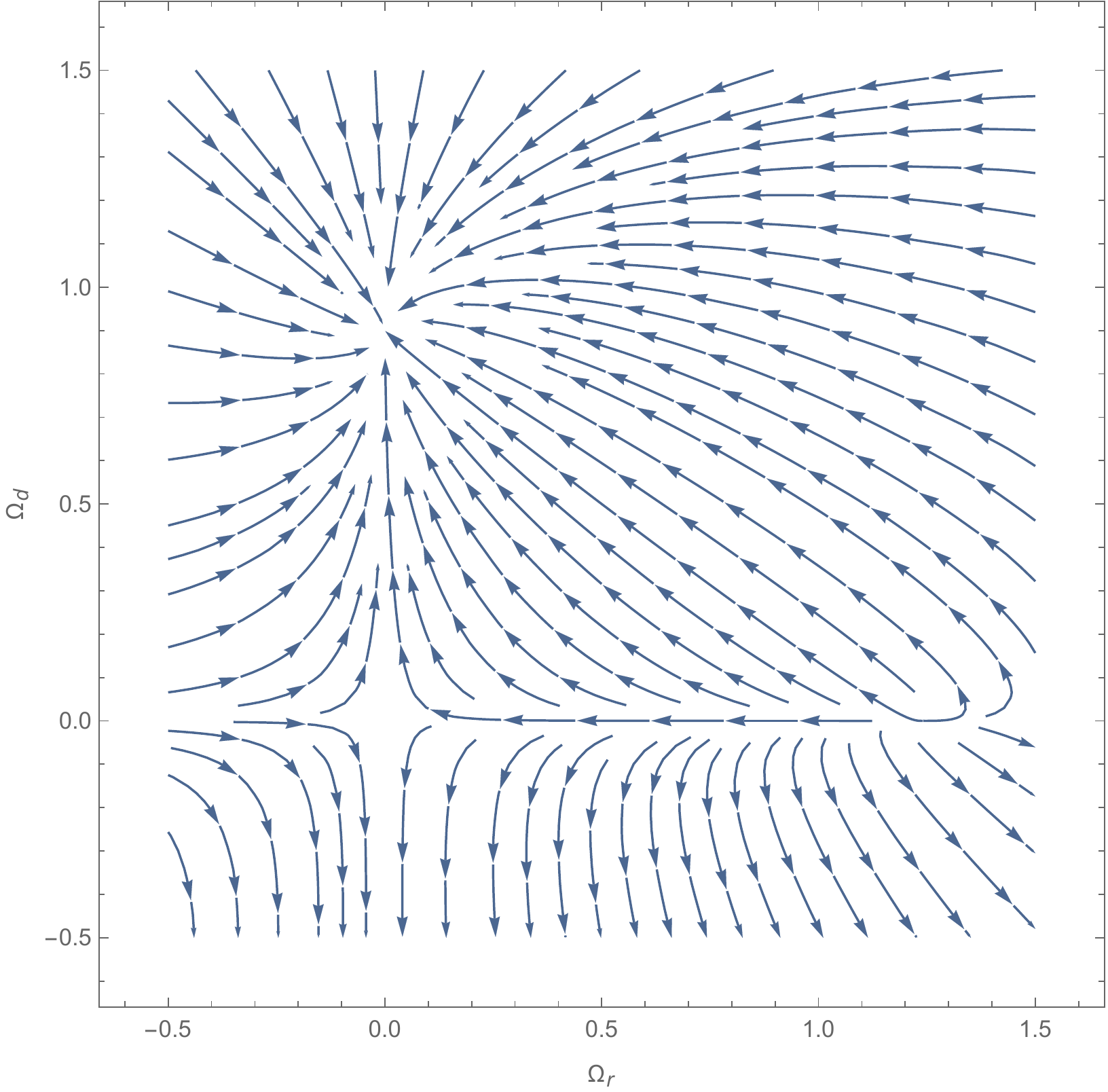}
	\caption{The phase space diagram of the system described by Eqs. (\ref{eq7}) and (\ref{eq8}) for $c_1=0.1$ and $c_2=0.01$. The radiation dominated fixed point $(\Omega_r,\ \Omega_d)=( 3 c_1-6 c_2+1,\ 0)$ is unstable, the matter dominated one $(\Omega_r,\ \Omega_d)=(0,\ 0)$, is a saddle fixed point and the dark energy dominated universe $(\Omega_r,\ \Omega_d)=(0,\ 1-c_1)$ is a stable fixed point.}
	\label{fig:figure 1}
\end{figure}

\begin{figure}[!htb]
	\minipage{0.31\textwidth}
	\includegraphics[width=\linewidth]{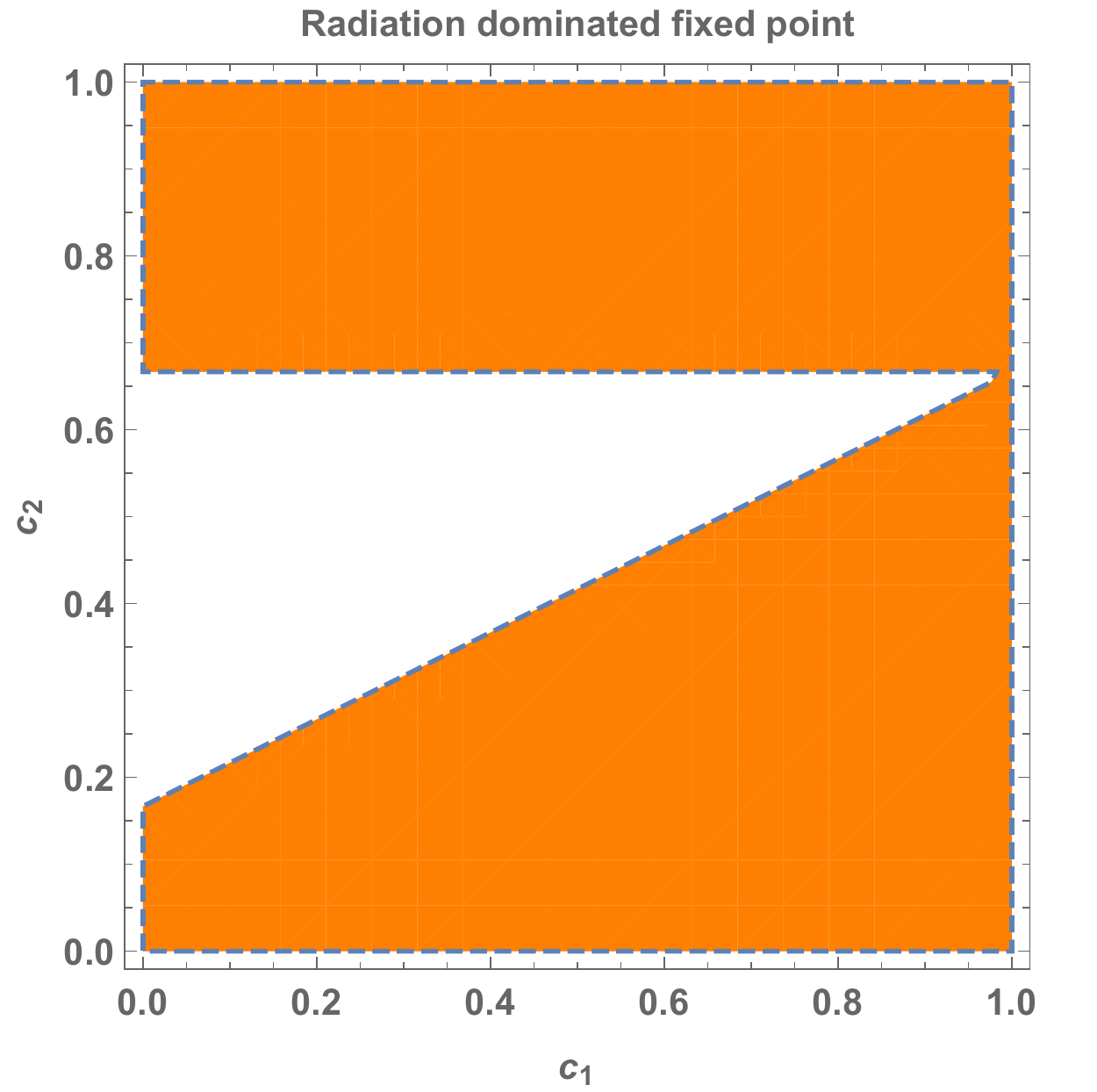}
	%\caption{Radiation}\label{fig:figure 7}
	\endminipage\hfill
	\minipage{0.31\textwidth}
	\includegraphics[width=\linewidth]{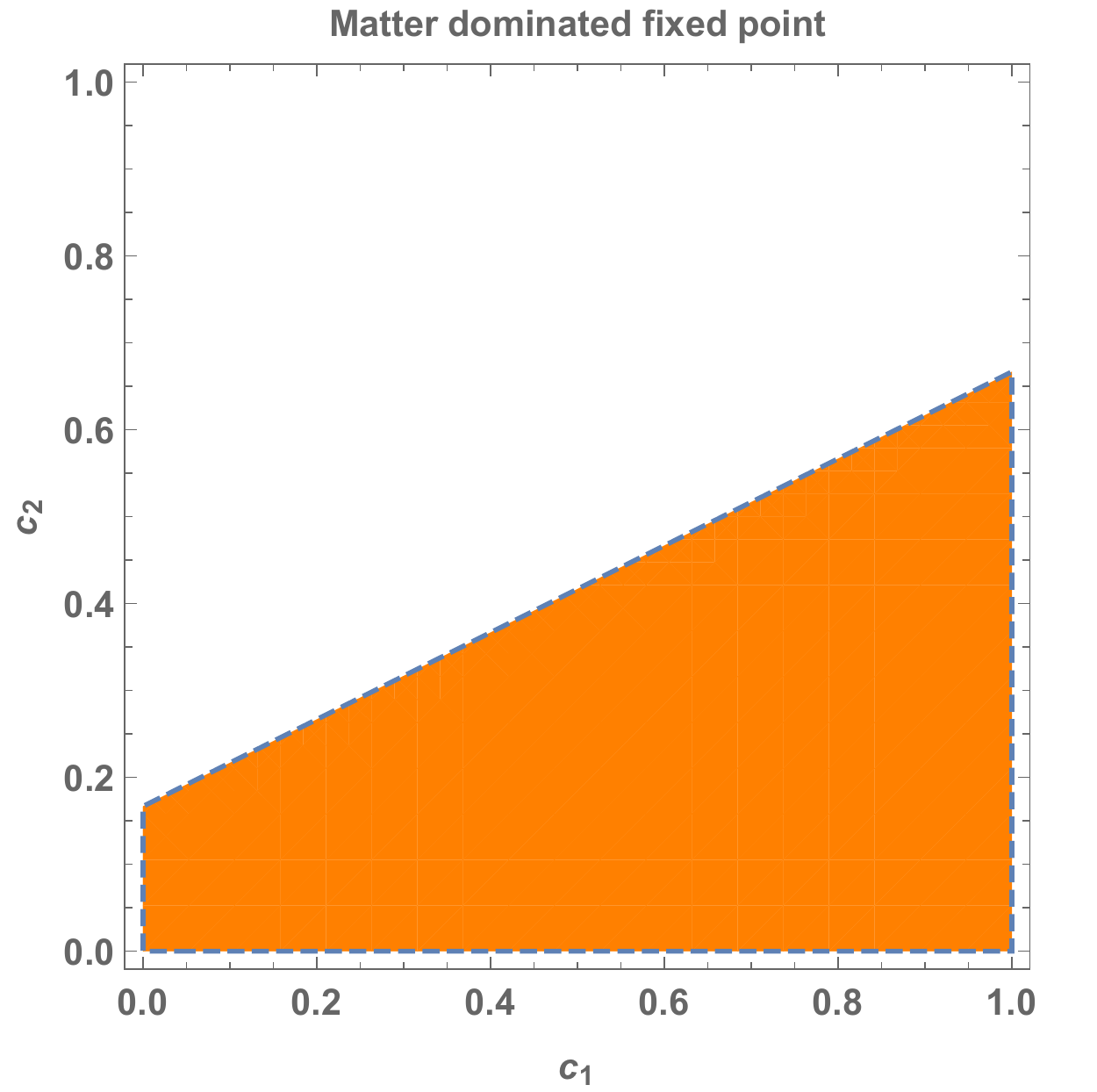}
	%	\caption{Matter}\label{fig:figure 8}
	\endminipage\hfill
	\minipage{0.31\textwidth}%
	\includegraphics[width=\linewidth]{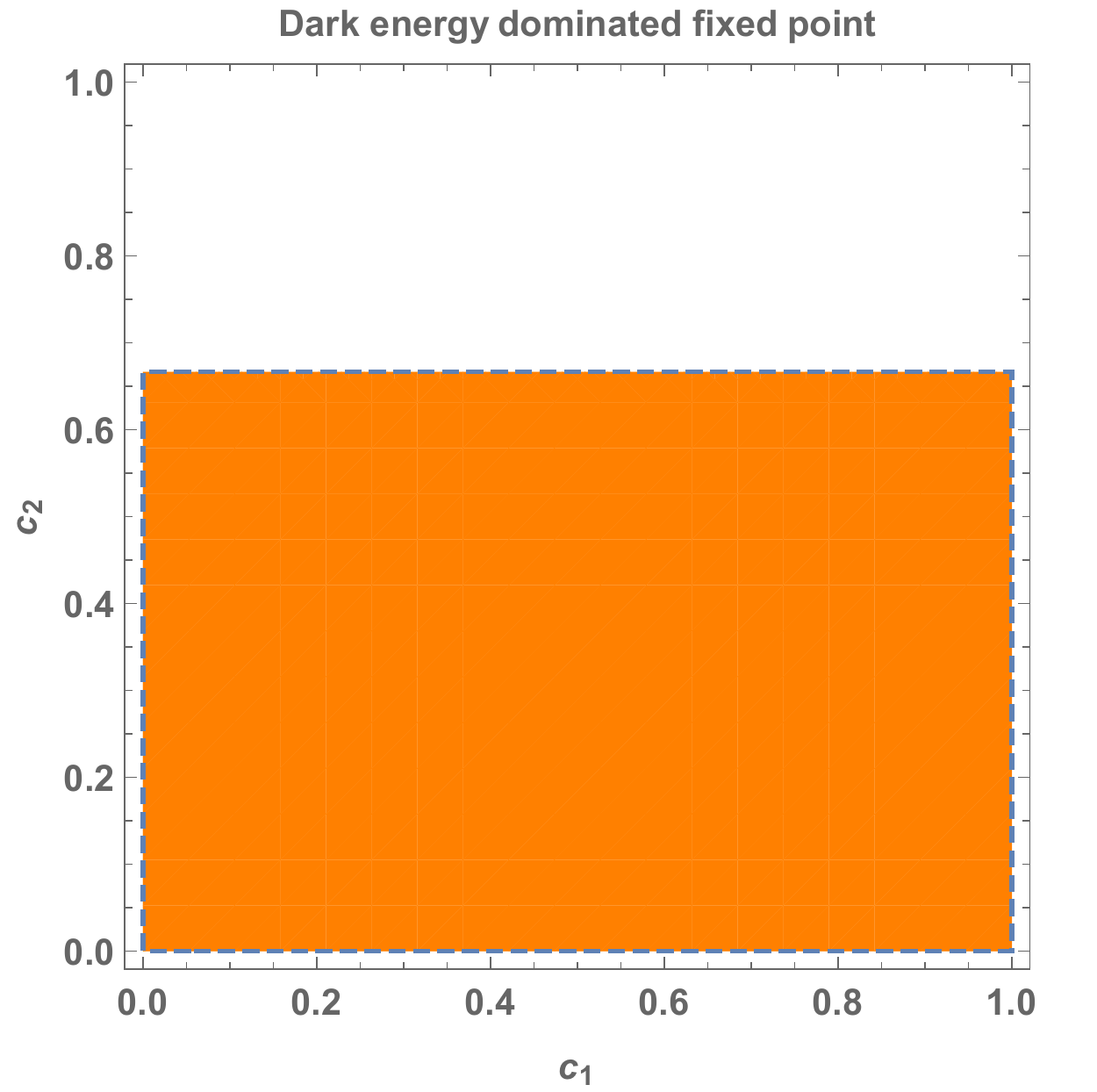}
	%	\caption{Dark energy}\label{fig:figure 9}
	\endminipage
	\caption{For the theory to be consistent with standard cosmology expectation the radiation, matter and dark energy dominated fixed points must be unstable, saddle and stable ones, respectively. The orange regions in above panels show the values of $c_1$ and $c_2$ that will satisfy this expectation for the model discussed in Sec.(\ref{Sec:2}) and by using the eigenvalues obtained in Eq. \eqref{eq12}.}\label{fig:figure 7}
\end{figure}

\begin{figure} % figure 7
	\center
	\includegraphics[scale=0.55]{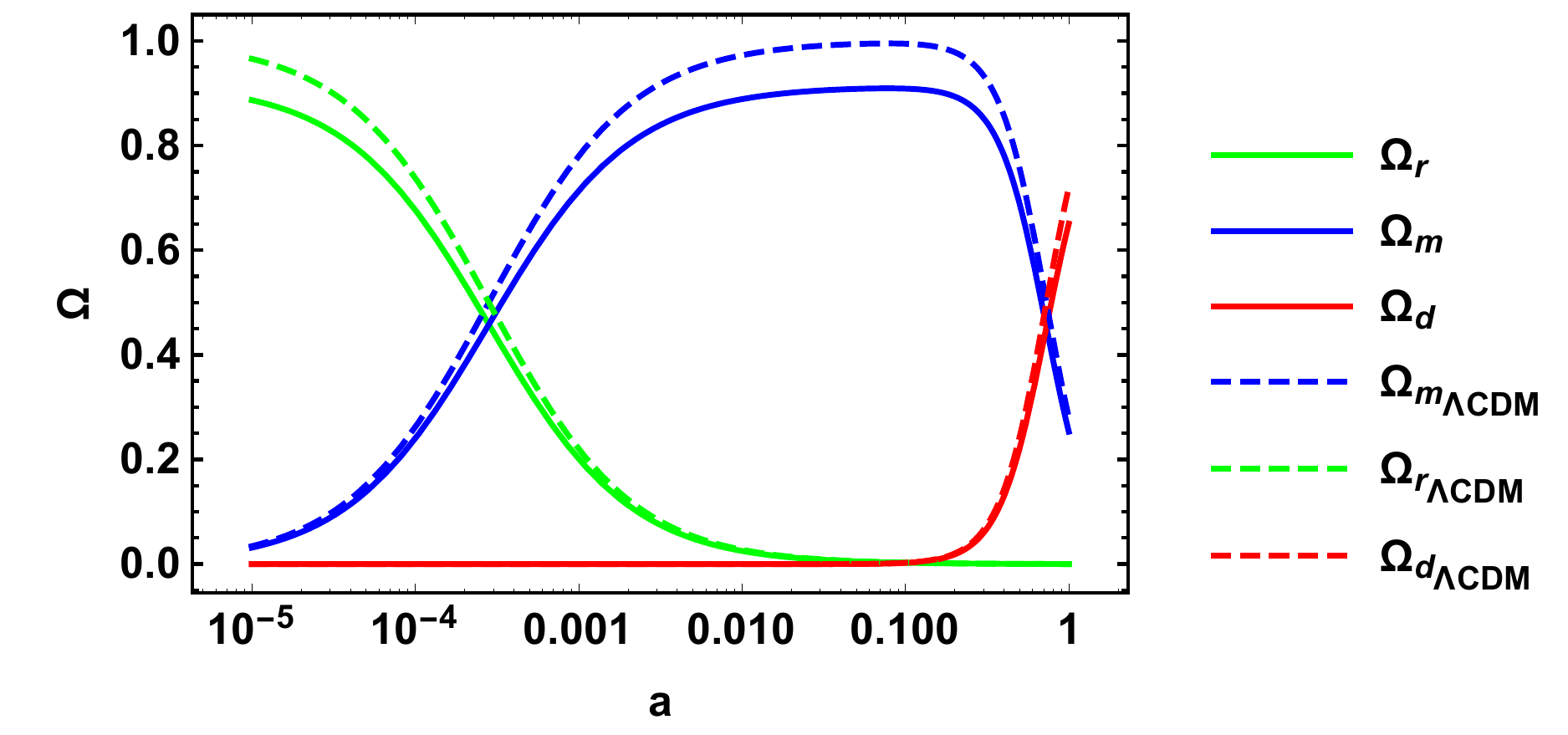}
	\caption{The evolution of energy portions of the universe with respect to scale factor for the model studied in Sec. (\ref{Sec:2}). Here we assume that $c_1=0.1$, $c_2=0.01$. $\Omega_{r0}=0.0004$, $\Omega_{m0}=0.2996$ and $\Omega_{d0}=0.7$ are the present values of radiation, matter and dark energy portions of the universe which have been used here.}
	\label{fig:figure 2}
\end{figure}

\begin{itemize}
	\item A fixed point is stable if both eigenvalues of the Jacobian matrix are negative.
	\item It is a saddle point if one of the eigenvalues is negative while the other is positive.
	\item Finally it is an unstable fixed point if both eigenvalues are positive.
\end{itemize}

Using Eqs.\eqref{eq9} and \eqref{eq11}, the eigenvalues of Jacobian matrix for the fixed points in Eq.\eqref{eq121} came out to be as follows:
\begin{eqnarray}
\label{eq12}
&&1.\ (\lambda_1=\frac{3 c_1+3 \sqrt{(c_1-1)^2}-12 c_2+5}{2-3 c_2},\ \lambda_2=\frac{3 c_1-3 \sqrt{(c_1-1)^2}-12 c_2+5}{2-3 c_2}),  \nonumber\\
&&2.\ (\lambda_1=\frac{-6 c_1+6 c_2+2 \sqrt{(3 c_2-2)^2}+2}{2-3 c_2},\ \lambda_2=\frac{2 \left(3 c_1-3 c_2+\sqrt{(3 c_2-2)^2}-1\right)}{3 c_2-2}),\\
&&3.\ (\lambda_1=-\frac{\sqrt{(3 c_1-6 c_2+1)^2}+3 c_1+6 c_2-7}{3 c_2-2},\ \lambda_2=\frac{\sqrt{(3 c_1-6 c_2+1)^2}-3 c_1-6 c_2+7}{3 c_2-2}).\nonumber
\end{eqnarray}

Applying different values of $c_1$ and $c_2$ (considering $0<c_1<1$ and $0<c_2<1$), we have found that for certain values of these parameters the first fixed point which represents the radiation dominated universe, has two positive eigenvalues and is an unstable point. The second one has a negative and a positive eigenvalue, thus it is a saddle point and represents the matter dominated universe. Finally dark energy epoch is a stable fixed point with two negative eigenvalues and attracts all the universe flux.

The observational evidence shows that the universe have had obvious time ordering for different epochs. There was a radiation dominated universe followed by a matter dominated universe. The last epoch is where dark energy or its alternatives has the main role. A successful gravitational theory must satisfy these stabilities and their ordering. Our results indicate that the theory that has been discussed here, is consistent with the standard cosmology expectations for certain values of $c_1$ and $c_2$ (Fig. \ref{fig:figure 7}). In this way we have restricted parameters $c_1$ and $c_2$ in the entropic force scenario. The phase space diagram and the evolution of energy \textcolor{black}{components} with respect to scale factor are depicted in Figs.(\ref{fig:figure 1}) and (\ref{fig:figure 2}).

%%%%%%%%%%%%%%%%%%%%%%%%%%%%%%%%%%%%%%%%%%%%%%%
\section{$\Omega_d=0$ and $\Omega_c=c_1+c_2\dfrac{\dot{H}}{H^2}$}
\label{Sec:3}
%%%%%%%%%%%%%%%%%%%%%%%%%%%%%%%%%%%%%%%%%%%%%%%%%%%%%%%%%%%%%%%%%%%%%
Here we would like to study the role of entropic force terms in dynamical behavior of the universe and in the absence of dark energy ($\Omega_d=0$). So it is natural to write the energy density corresponding to entropic force as $\rho_c=\frac{3}{k}(c_1H^2+c_2\dot{H})$, with the continuity equation $\dot{\rho}_c+3H(\rho_c+p_c)=0$. Thus Eqs. \eqref{eq5} and \eqref{eq6} turn into the following equations:
\begin{eqnarray}
\label{eq131}
&&1=\Omega_r+\Omega_m+\Omega_c,\\
\label{eq14}
&&\dfrac{\dot{H}}{H^2}=\dfrac{-1}{2}\Omega_m-\Omega_r+\Omega_c-1,
\end{eqnarray}
\noindent where $\Omega_c$ is the entropic force energy portion and is defined as $\Omega_c=c_1+c_2\dfrac{\dot{H}}{H^2}$. Therefore $\Omega_r$, $\Omega_m$ and $\Omega_c$ are the energy \textcolor{black}{components} of the universe and one of them will be omitted due to Eq.\eqref{eq131}. Here we will omit $\Omega_c$ out of the three, thus the derivatives of the two remaining ones with respect to the number of e-foldings are:
\begin{eqnarray}
\label{eq15}
&&\Omega'_r=-4\Omega_r-2\Omega_r \dfrac{\dot{H}}{H^2},\\
&&\Omega'_m=-3\Omega_m-2\Omega_m\dfrac{\dot{H}}{H^2}.
\end{eqnarray}

\textcolor{black}{Solving these two equations together (i.e. $\Omega'_r=0$ and $\Omega'_m=0$) to obtain the fixed points, and calculating the eigenvalues of Jacobian matrix for each of them, we have found the following fixed points and their corresponding eigenvalues ($\lambda_1$ and $\lambda_2$):}
\begin{eqnarray}
\label{eq16}
&&1.\ (\Omega_r,\ \Omega_m)=(1,\ 0)\Longrightarrow(\lambda_1=1,\ \lambda_2=4),\nonumber\\
&&2.\ (\Omega_r,\ \Omega_m)=(0,\ 1)\Longrightarrow(\lambda_1=-1,\ \lambda_2=3),\nonumber\\
&&3.\ (\Omega_r,\ \Omega_m)=(0,\ 0)\Longrightarrow(\lambda_1=-4,\ \lambda_2=-3).
\end{eqnarray}

\begin{figure} % figure 3
	\center
	\includegraphics[scale=0.5]{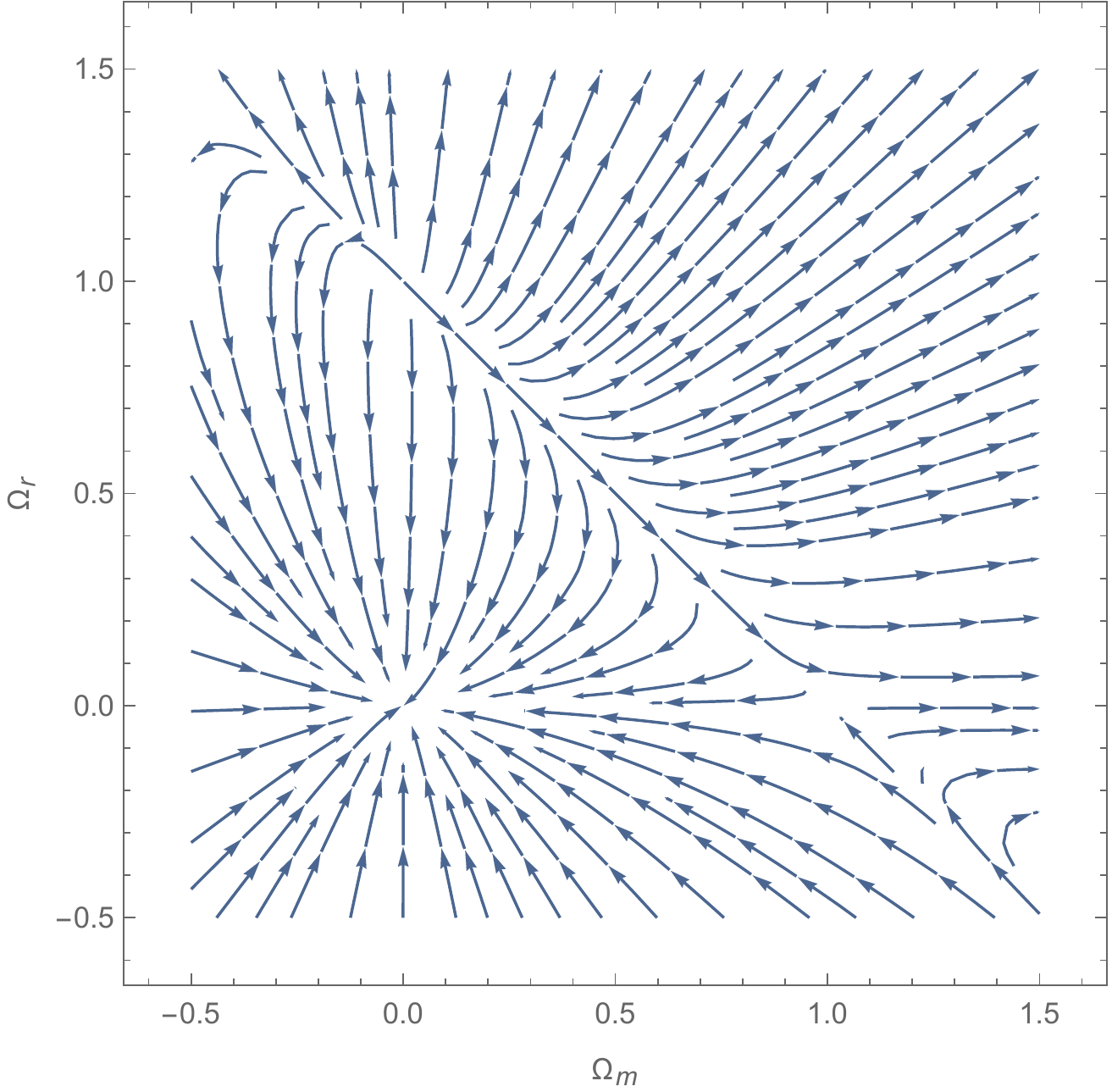}
	\caption{The phase space diagram of the system described by Eqs. (\ref{eq131}) and (\ref{eq14}). The radiation dominated fixed point $(\Omega_r,\ \Omega_m)=(1,\ 0)$ is unstable, the matter dominated one $(\Omega_r,\ \Omega_m)=(0,\ 1)$, is a saddle fixed point and the entropic force dominated universe $(\Omega_r,\ \Omega_m)=(0,\ 0)$ is a stable fixed point.}
	\label{fig:figure 3}
\end{figure}

Thus it is concluded that the radiation dominated epoch (the first fixed point) is an unstable fixed point and the universe flux lines leave this point toward the next one. The second fixed point which represents the matter dominated universe, is a saddle point, so the flux lines are attracted to this point from one direction and leaves it from another one. Thus there would be a third epoch to attract all the flux lines. It is where the entropic force energy density has a lead role in the universe (the third fixed point). As the eigenvalues of Jacobian matrix are negative for this point, it is a stable fixed point and a final attractor. Therefore using entropic force effect, it is possible to describe the late time expansion of the universe without using the concept of dark energy. Our result is presented in Fig.(\ref{fig:figure 3}).
%%%%%%%%%%%%%%%%%%%%%%%%%%%%%%%%%%%%%%%%%%%%%%%%%%%%%%
\section{The case with both $\Omega_d$ and $\Omega_c$}
\label{Sec:4}
%%%%%%%%%%%%%%%%%%%%%%%%%%%%%%%%%%%%%%%%%%%%%%%%%%%%%%%%%
Here, let us consider Eqs. \eqref{eq1} and \eqref{eq2} by assuming that $\Omega_c=c_1+c_2\dfrac{\dot{H}}{H^2}$. Thus we will have a universe in which the entropic force and dark energy are both assumed to be energy \textcolor{black}{components} of the universe. Using these definitions Eqs. \eqref{eq1} and \eqref{eq2} will be given as follows:
\begin{eqnarray}
\label{eq17}
&&1=\Omega_r+\Omega_m+\Omega_d+\Omega_c,\\
\label{eq18}
&&\dfrac{\dot{H}}{H^2}=\Omega_c+\Omega_d-\dfrac{1}{2}\Omega_m-\Omega_r-1.
\end{eqnarray}

Where due to Eq.\eqref{eq17}, the system will be reduced from four dimensional system to a three dimensional one (Here we have chosen $\Omega_c$ to be omitted between the four).  Thus the derivative of three remaining $\Omega$ with respect to number of e-foldings will be obtained:
\begin{eqnarray}
\label{eq19}
\Omega'_m&=&-2 \Omega_m \left(-1 + \Omega_d - \frac{\Omega_m}{2} + (1 - \Omega_d - \Omega_m - \Omega_r) - \Omega_r\right)-3 \Omega_m,\\
\Omega'_r&=&-2 \Omega_r \left(-1 + \Omega_d - \frac{\Omega_m}{2} + (1 - \Omega_d - \Omega_m - \Omega_r) - \Omega_r \right)-4 \Omega_r,\\
\Omega'_d&=&-2 \Omega_d \left(-1 + \Omega_d - \frac{\Omega_m}{2} + (1 - \Omega_d - \Omega_m - \Omega_r) - \Omega_r\right).
\end{eqnarray}
\begin{figure} % figure 4
	\center
	\includegraphics[scale=0.5]{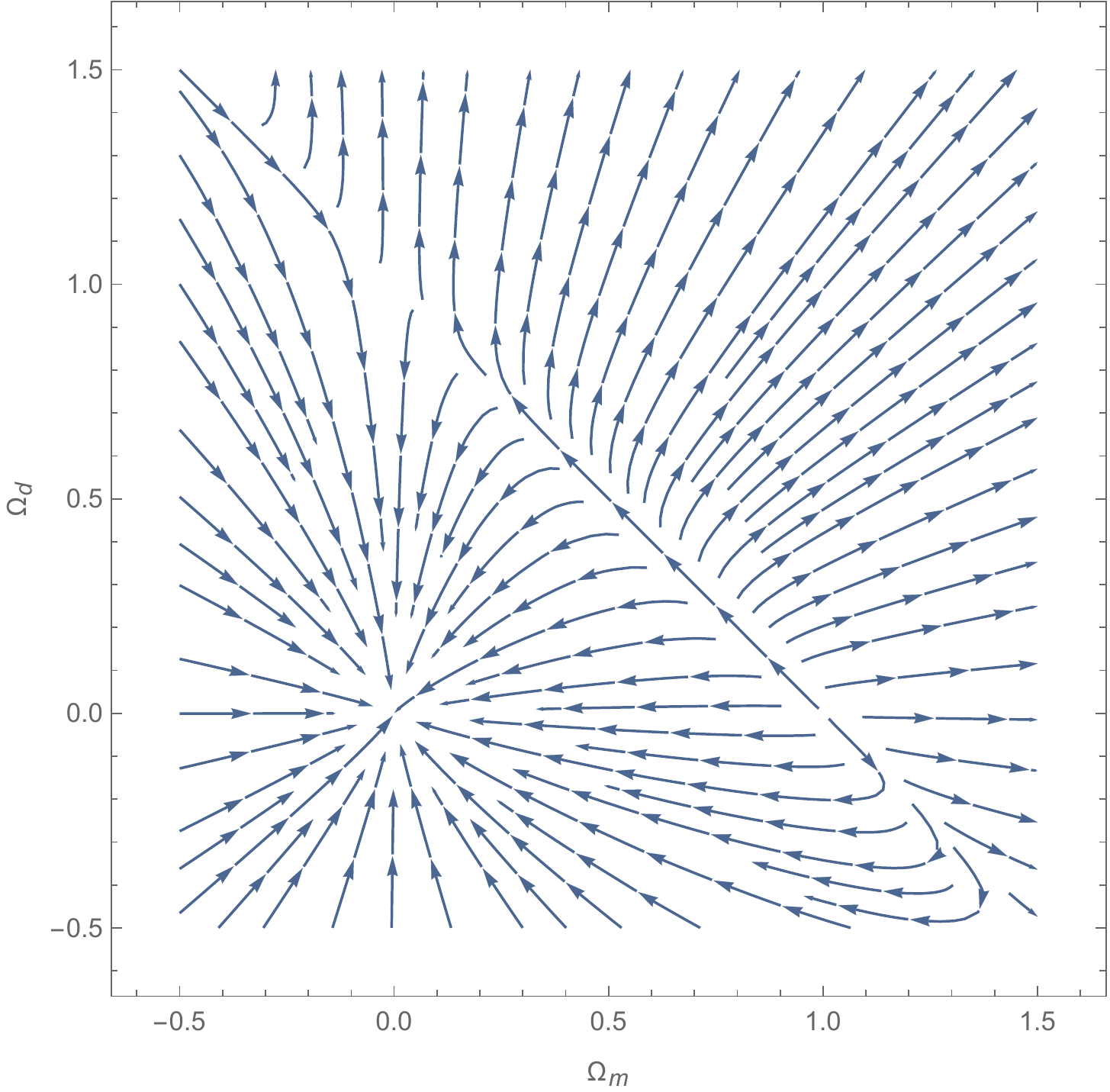}
	\caption{The phase space diagram of the system described by Eqs. (\ref{eq17}) and (\ref{eq18}) with $\Omega_r=0$. According to Eq.\eqref{eq20}, the radiation dominated fixed point $(\Omega_r,\ \Omega_m,\ \Omega_d)=(1,\ 0,\ 0)$ is unstable, the matter dominated one $(\Omega_r,\ \Omega_m,\ \Omega_d)=(0,\ 1,\ 0)$, is a saddle fixed point, and the dark energy universe $(\Omega_r,\ \Omega_m,\ \Omega_d)=(0,\ 0,\ 1)$ is also a saddle fixed point. Finally the entropic force dominated universe $(\Omega_r,\ \Omega_m,\ \Omega_d)=(0,\ 0,\ 0)$ is a stable fixed point. }
	\label{fig:figure 4}
\end{figure}

As this is a three dimensional dynamical system, solving these three equations together (i.e. $\Omega'_r=0,\ \Omega'_m=0,\ \Omega'_d=0$), gives us four fixed points where each of them will have three \textcolor{black}{corresponding} Jacobian matrix eigenvalues as bellow:
\begin{eqnarray}
\label{eq20}
&&1.\ (\Omega_r,\ \Omega_m,\ \Omega_d)=(1,\ 0,\ 0)\Longrightarrow(\lambda_1>0 ,\ \lambda_2>0 ,\ \lambda_3>0),\nonumber\\
&&2.\ (\Omega_r,\ \Omega_m,\ \Omega_d)=(0,\ 1,\ 0)\Longrightarrow(\lambda_1>0 ,\ \lambda_2>0 ,\ \lambda_3<0),\nonumber\\
&&3.\ (\Omega_r,\ \Omega_m,\ \Omega_d)=(0,\ 0,\ 1)\Longrightarrow(\lambda_1<0 ,\ \lambda_2<0 ,\ \lambda_3>0),\nonumber\\
&&4.\ (\Omega_r,\ \Omega_m,\ \Omega_d)=(0,\ 0,\ 0)\Longrightarrow(\lambda_1<0 ,\ \lambda_2<0 ,\ \lambda_3<0).
\end{eqnarray}

These results implies that the radiation dominated epoch (the first fixed point) is an unstable fixed point while the matter dominated epoch (the second one) represents a saddle fixed point. The dark energy universe is a saddle fixed point (the third one) and will be followed by a stable entropic force dominated universe (the forth fixed point).  

\section{Conclusion}
In this paper we have studied the dynamical behavior of the universe considering the effect of entropic force (entropic acceleration) on the equations of motion. Entropic force leads to additional terms in Friedmann equations which may have a significant role in the dynamical behavior of the universe. \\
\textcolor{black}{An important result of considering the entrpoic force scenario is that, there is no need to take into account the dark energy to explain the late time acceleration of the universe.}

Using Eq. \eqref{eq01}, and assuming the energy \textcolor{black}{components} of the universe to be $\rho_r,\ \rho_m$ and $\rho_d$, it is came out that there are three fixed points for the model which are radiation, matter and dark energy dominated universe and according to the eigenvalues of Jacobian matrix, the stability of them depends on the values of $c_1$ and $c_2$. For the theory to be consistent with standard cosmology expectation $c_1$ and $c_2$ must be in the intervals that was presented in Fig. (\ref{fig:figure 7}).\textcolor{black}{ These intervals are consistent with the values that Easson, Frampton and Smoot have applied in their calculation \cite{1002.4278}.} The proper values of $c_1$ and $c_2$ shows that there would be a radiation dominated epoch in the universe followed by a matter dominant one and finally our universe will witness a dark energy dominated epoch. \textcolor{black}{Thus we have put a boundary on the values of $c_1$ and $c_2$ by using the dynamical method calculations and these boundaries are consistent with the previous works \cite{1002.4278,1309.7827}.}

Entropic force universe is introduced as an alternative to dark energy, thus we investigated the dynamics of this theory in the absence of dark energy. In this case the energy \textcolor{black}{components} of universe are radiation, matter and entropic force, $\Omega_c$, as in Sec. (\ref{Sec:3}).\textcolor{black}{ The fixed points of this case and their stability represents that the radiation and matter dominated epochs are unstable and saddle fixed points of this case, respectively, while the entropic force will have the lead role in the late time of the universe as a stable fixed point. It implies that} \textcolor{black}{ the entropic force can explain the late time acceleration of the universe instead of dark energy.}

Also considering both dark energy and entropic force to be the energy \textcolor{black}{components} of the universe, we have found an interesting result. Again the radiation and matter dominated fixed points are unstable and saddle fixed points, the dark energy epoch will be a saddle fixed point of the model in Sec. (\ref{Sec:4}) and the entropic force dominated fixed point is a stable one. \textcolor{black}{Thus there would be a radiation dominated universe followed by a matter dominant epoch, prior to a dark energy dominated universe and the last epoch will be an entropic force dominated epoch.}

\bibliographystyle{ws-ijmpd}
\bibliography{EF}

\begin{thebibliography}{10}

\bibitem{9812133}
S.~Perlmutter {\em et~al.}, {\em Astrophys. J.} {\bf 517}  (1998)   565,
  \href{http://arxiv.org/abs/9812133}{{\ttfamily arXiv:9812133 [astro-ph]}}.

\bibitem{9805201}
A.~Reiss {\em et~al.}, {\em Astron. J.} {\bf 116}  (1998)   1009,
  \href{http://arxiv.org/abs/9805201}{{\ttfamily arXiv:9805201 [astro-ph]}}.

\bibitem{weinberg}
S.~Weinberg, {\em Rev. Mod. Phys} {\bf 61}  (1989) 1.

\bibitem{Husain}
V.~Husain and B.~Qureshi, {\em Phys. Rev. Lett.} {\bf 116}  (2016)   061302,
  \href{http://arxiv.org/abs/1508.07664}{{\ttfamily arXiv:1508.07664 [gr-qc]}}.

\bibitem{0803.0982}
J.~Frieman, M.~Turner and D.~Huterer, {\em Ann.Rev.Astron.Astrophys.} {\bf 46}
  (2008) 385, \href{http://arxiv.org/abs/0803.0982}{{\ttfamily arXiv:0803.0982
  [astro-ph]}}.

\bibitem{0705.1032}
S.~Tsujikawa, {\em Phys.Rev.D} {\bf 76}  (2007)   023514,
  \href{http://arxiv.org/abs/0705.1032}{{\ttfamily arXiv:0705.1032
  [astro-ph]}}.

\bibitem{0805.1726}
T.~P. Sotiriou and V.~Faraoni, {\em Rev. Mod. Phys.} {\bf 82}  (2010) 451,
  \href{http://arxiv.org/abs/0805.1726}{{\ttfamily arXiv:0805.1726 [gr-qc]}}.

\bibitem{1104.2669}
T.~Harko, F.~S. Lobo, S.~Nojiri and S.~D. Odintsov, {\em Phys. Rev. D} {\bf 84}
   (2011)   024020, \href{http://arxiv.org/abs/1104.2669}{{\ttfamily
  arXiv:1104.2669 [gr-qc]}}.

\bibitem{verlinde}
E.~P. Verlinde, {\em JHEP} {\bf 11}  (2011)   029,
  \href{http://arxiv.org/abs/1001.0785}{{\ttfamily arXiv:1001.0785 [hep-th]}}.

\bibitem{unrah}
W.~Unruh, {\em Phys. Rev. D} {\bf 14}  (1976)   870.

\bibitem{1002.4278}
D.~A. Easson1, P.~H. Frampton1 and G.~F. Smoot, {\em Phys. Lett. B} {\bf 696}
  (2011) 273, \href{http://arxiv.org/abs/1002.4278}{{\ttfamily arXiv:1002.4278
  [hep-th]}}.

\bibitem{1003.1528}
D.~A. Easson1, P.~H. Frampton1 and G.~F. Smoot, {\em Int. J. Mod. Phys. A} {\bf
  27}  (2012)   1250066, \href{http://arxiv.org/abs/1003.1528}{{\ttfamily
  arXiv:1003.1528 [hep-th]}}.

\bibitem{1003.4526}
Y.-F. Cai, J.~Liu and H.~Li, {\em Phys. Lett. B} {\bf 690}  (2010) 213,
  \href{http://arxiv.org/abs/1003.4526}{{\ttfamily arXiv:1003.4526
  [astro-ph.CO]}}.

\bibitem{1005.0790}
R.~Casadio and A.~Gruppuso, {\em Phys. Rev. D} {\bf 84}  (2011)   023503,
  \href{http://arxiv.org/abs/1005.0790}{{\ttfamily arXiv:1005.0790 [gr-qc]}}.

\bibitem{1005.1445}
H.~Wei, {\em Phys. Lett. B} {\bf 692}  (2010) 167,
  \href{http://arxiv.org/abs/1005.1445}{{\ttfamily arXiv:1005.1445 [gr-qc]}}.

\bibitem{1005.2240}
Y.~S. Myung, {\em Astrophys. Space Sci.} {\bf 335}  (2011) 553,
  \href{http://arxiv.org/abs/1005.2240}{{\ttfamily arXiv:1005.2240 [hep-th]}}.

\bibitem{1309.7827}
M.~A. Abchouyeh, B.~Mirza and Z.~Sherkatghanad, {\em Gen. Relativ. Gravit.}
  {\bf 46}  (2014)   1617, \href{http://arxiv.org/abs/1309.7827}{{\ttfamily
  arXiv:1309.7827 [gr-qc]}}.

\bibitem{1409.5585}
C.~G. Boehmer and N.~Chan, {\em LTCC Advanced Mathematics Series} {\bf 5}
  (2017) 121, \href{http://arxiv.org/abs/1409.5585}{{\ttfamily arXiv:1409.5585
  [gr-qc]}}.

\bibitem{1512.09281}
R.~An, X.~Xu, B.~Wang and Y.~Gong, {\em Phys. Rev. D} {\bf 93}  (2016)
  103505, \href{http://arxiv.org/abs/1512.09281}{{\ttfamily arXiv:1512.09281
  [gr-qc]}}.

\end{thebibliography}
\end{document}